\documentclass[12pt]{amsart}
\usepackage[utf8]{inputenc}
\usepackage[english]{babel}

\usepackage{amssymb}
\usepackage{amsmath}
\usepackage{amsthm}
\usepackage{leftidx}
\usepackage{mathdots}
\usepackage{amssymb}
\usepackage{mathrsfs}

\usepackage{epsfig}
\usepackage{verbatim}
\newcommand{\lra}{{\longrightarrow}}
\newcommand{\eproof}{\hfill\rule{2.2mm}{3.0mm}}

\newcommand{\Proof}{\noindent {\bf Proof.~~}}

\newcommand{\R}{{\mathbb R}}

\newcommand{\C}{{\mathbb C}}

\newcommand{\PP}{{\mathbb P}}

\renewcommand{\eqref}[1]{(\ref{#1})}

\newcommand{\innerp}[1]{\langle #1 \rangle}

\newcommand{\mhsp}{\hspace{2em}}
\newcommand{\abs}[1]{\lvert#1\rvert}

\newcommand{\A}{{\mathcal A}}

\newcommand{\rank}{{\rm rank}}

\newcommand{\Tr}{{\rm Tr}}

\newcommand{\vx}{{\mathbf x}}
\newcommand{\vy}{{\mathbf y}}

\newcommand{\vw}{{\mathbf w}}

\newcommand{\vv}{{\mathbf v}}

\newcommand{\vf}{{\mathbf f}}

\renewcommand{\H}{{\mathbb F}}

\newcommand{\M}{{\mathcal M}}
\newcommand{\NN}{{\mathcal N}}

\newcommand{\LL}{\mathbf L}

\newcommand{\G}{{\mathcal G}}

\newtheorem{prop}{Proposition}[section]
\newtheorem{lem}[prop]{Lemma}
\newtheorem{defi}{Definition}[section]
\newtheorem{coro}[prop]{Corollary}
\newtheorem{theo}[prop]{Theorem}
\newtheorem{remark}[prop]{Remark}

\begin{document}
\baselineskip 18pt
\title[Almost Everywhere Injectivity Conditions for  Matrix Recovery]{Almost everywhere injectivity conditions for  the matrix recovery problem}
\author{Yi Rong}
\address{Department of Mathematics  \\ Hong Kong University of Science and Technology\\
Clear Water Bay, Kowloon, Hong Kong}
\email{yrong@ust.hk}
\author{Yang Wang}
\thanks{Yang Wang was supported in part by
 the NSFC grant 91630203,
 the Hong Kong Research Grant Council grant 16306415 and 16317416.
      Zhiqiang Xu was supported  by NSFC grant (91630203, 11688101),
Beijing Natural Science Foundation (Z180002).}
\address{Department of Mathematics  \\ Hong Kong University of Science and Technology\\
Clear Water Bay, Kowloon, Hong Kong}
\email{yangwang@ust.hk}
\author{Zhiqiang Xu}
\address{LSEC, Inst.~Comp.~Math., Academy of
Mathematics and System Science,  Chinese Academy of Sciences, Beijing, 100091, China \newline
School of Mathematical Sciences, University of Chinese Academy of Sciences, Beijing 100049, China}
\email{xuzq@lsec.cc.ac.cn}

\begin{abstract}
The aim of matrix recovery is to recover $P\in \M\subset \H^{p\times q}$ from
$
 \LL_\A(P) = (\Tr(A_1^T P), \Tr(A_2^T P), \dots, \Tr(A_N^T P))^T
$
with $A_j\in V_j\subset \H^{p\times q}$,
which is raised in many   areas. In this paper, we build up a framework for almost everywhere  matrix recovery which means $\LL_\A$ is almost everywhere injectivity on $\M$. We mainly focus on the following question: how many measurements are needed to recover almost all the  matrices in $\M$?
For the case where both $\M$ and $V_j$ are algebraic varieties, we use the tools from algebraic geometry to study the question and present some results to address  it under many different settings.

\end{abstract}
\maketitle

\section{Introduction}
\setcounter{equation}{0}

\subsection{Matrix Recovery Problems }

The matrix recovery problem has gained much attention in recent years. The general formulation of the problem is that there is an unknown  $p\times q$ real or complex matrix $P$ and we would like to recover the matrix $P$ from a collection of  measurements or samples. A typical such problem is the so-called {\em Netflix Problem}, where we know the value of some but not all entries, and the matrix in question has low rank. The aim is to fully recover the matrix from the partial set of entries. The Netflix Problem has seen extensive study because of its broad applications in many other areas (see \cite{lowrank1,lowrank2,lowrank3}).

The Netflix Problem is a special case of matrix recovery from linear measurements, which can be phrased generally as follows: For $1 \leq j \leq N$ let $L_j: \H^{p\times q} \lra \H$ be linear maps, where $\H =\R$ or $\C$. Given $L_1(P),\ldots,L_N(P)$, can we recover $P\in \H^{p\times q}$ ? The ability to recover $P$ depends on the properties of $P$ and $L_j$, and we also need to have {\em enough} measurements.

It is well known that a linear map $L: \H^{p\times q} \lra \H$ can always be represented in the form of $L(P) = \Tr(A^TP)$ for  $A \in \H^{p\times q} $, and such a representation is one-to-one. Now let
$\A=(A_j)_{j=1}^N$ be a sequence of matrices with each $A_j \in \H^{p\times q}$. We denote by $\LL_\A: \H^{p\times q} \lra \H^N$  the map given by
\begin{equation}  \label{eq:trace-map}
        \LL_\A(P) = (\Tr(A_1^T P), \Tr(A_2^T P), \dots, \Tr(A_N^T P))^T, \mhsp P\in \H^{p\times q}.
\end{equation}
Matrix recovery problems aim to recover a matrix in a subset of $\H^{p\times q}$ from linear measurements. Let $\M \subseteq \H^{p\times q}$ be the subset of interest. We say  $\A=(A_j)_{j=1}^N$ where $A_j \in \H^{p\times q}$ has the {\em $\M$-recovery property} if every $P\in \M$ is uniquely determined by $\LL_\A(P)$. In other words, the map $\LL_\A$ is injective on $\M$.

One particular class of matrices of interest is the set of all rank $r$ or less matrices, which we denote by
\begin{equation}  \label{eq:rankr}
  {\mathcal M}_{p\times q,r}(\H):=\Bigl\{Q\in \H^{p\times q}: {\rank}(Q)\leq r\Bigr\},
  \mhsp \H=\R~{\rm or}~\C.
\end{equation}
For example, it is known that at least $N \geq 4rn -4r^2$ linear measurements are needed to completely recover any  $P\in {\mathcal M}_{n\times n,r}(\C) $ where $0<r\leq n/2$, and furthermore $N \geq 4rn -4r^2$ linear measurements will suffice  (see \cite{Xu15,uniq}).

\subsection{Almost Everywhere Matrix Recovery}

As said before,  by providing enough measurements,
 we can recover all matrices in $\M_{n\times n,r}(\C)$, e.g. $N \geq 4rn-4r^2$ random measurements.   Sometimes we may have fewer measurements. Numerical experiments show that
it is possible  to recover most of matrices in $\M_{n\times n, r}(\C)$ from $N < 4rn-4r^2$ random measurements. So, sometimes,  even though we can not be able to recover    all matrices in a subset $\M$, we may be able to recover most of them nevertheless. The aim of this paper is to present conditions under which $\LL_\A$ is almost everywhere injective on $\M$.
This leads to the notion of {\em almost everywhere matrix recovery}.
\begin{defi}  \label{defi-Matrix-ae}{\rm
    Let $\M\subset \H^{p\times q}$ and $\A=(A_j)_{j=1}^N \in \left(\H^{p\times q}\right)^N$. We say $\LL_\A$ or simply $\A$ has the {\it $\M$-recovery property} if $\LL_\A$ is injective on $\M$. It has the {\it almost everywhere $\M$-recovery property} if for almost every $P\in \M$, we have $\LL_\A^{-1}\bigl(\LL_\A(P)\bigr) \cap \M= \{P\}$ where $\LL_\A^{-1}\bigl(\LL_\A(P)\bigr):=\{Q\in \H^{p\times q}: \LL_\A(Q)=\LL_\A(P)\}$.
}
\end{defi}
Here the easiest way to define ``almost everywhere'' is through the Hausdorff measure on $\M$. But since our study only focuses  on $\M$ that are ``nice'' such as algebraic varieties there should be no ambiguity. For the case where $\M$ is an algebraic variety, to show $\A$ has the almost everywhere $\M$-recovery property, it is enough to prove that there exists a subvariety $Y\subset \M$ with $\dim (Y)<\dim (\M)$ so that $\LL_\A^{-1}\bigl(\LL_\A(P)\bigr) \cap \M= \{P\}$ for any $P\in \M\setminus Y$.
 We say $\A=(A_j)_{j=1}^N \in \left(\H^{p\times q}\right)^N$ or $\LL_\A$ has the {\em almost everywhere rank $r$ matrix recovery property} if it has the almost everywhere $ {\mathcal M}_{p\times q,r}(\H)$-recovery property. Note that in this case $ {\mathcal M}_{p\times q,r}(\H)$ is an algebraic variety of dimension $r(p+q)-r^2$ (see \cite{alge}).

This paper studies the following questions: Let $\A=(A_j)_{j=1}^N \in \left(\H^{p\times q}\right)^N$. {\em What is the minimal measurement number  $N$ needed for $\LL_\A$ to have the almost everywhere rank $r$ matrix recovery property? } Or more generally, for a given subset $\M\subset \H^{p\times q}$, what is the minimal measurement number  $N$ needed for $\LL_\A$ to have the almost everywhere $\M$-recovery property? Note that in general we also have additional constraints on measurement matrices $\A$. The aim of  this paper is to present a series of results addressing these questions.
\subsection{Related Work}
In the context of matrix recovery, one already presents many conditions under which  $\A=(A_j)_{j=1}^N$  has $\M_{n\times n, r}(\H)$-recovery property \cite{uniq,uniq1,uniq2,uniq3}. In \cite{uniq}, it is proved that if $N\geq 4nr-4r^2$ and $A_1,\ldots,A_N$ are Gaussian random matrices, then $\A$ has $\M_{n\times n, r}(\H)$-recovery property with probability 1.  In \cite{uniq}, Eldar, Needell and  Plan conjecture   the measurement number $4nr-4r^2$ is tight. In \cite{Xu15}, Xu confirm the conjecture for the case $\H=\C$ and also disprove it for $\H=\R$.
In \cite{Bilinear}, Kech and Krahmer study the optimal injectivity conditions for bilinear inverse problems with employing the dimension theory in algebraic geometry. Particularly, assume that the the sparsity of the input vectors are $s_1$ and $s_2$ and  they show that if the measurement number $N\geq 2(s_1+s_2)-2$, almost all bilinear maps are injective on the set
of pairs of sparse vectors .

Under the setting of  $\M=\{\vx\vx^*:\vx\in \H^n\}\subset \H^{n\times n}$ and $A_j=\vf_j\vf_j^*$ with $\vf_j\in \H^n, j=1,\ldots,N$, $\A=(A_j)_{j=1}^N$ has the almost everywhere  $\M$-recovery property if and only if $(\vf_j)_{j=1}^N$ has the almost  phase retrieval  property. Here, we say
$(\vf_j)_{j=1}^N$ has the almost  phase retrieval  property if
one can recover almost all the $\vx\in \H^n$ (up to an unimodular constant) from $\abs{\innerp{\vf_j,\vx}}, j=1,\ldots,N$.
It is an active topic to present the smallest $N$ for which $(\vf_j)_{j=1}^N$  having the almost  phase retrieval property \cite{BCE06,FMW14,FSC05,Mix14}. For the case where $\H=\R$, it is known that $N\geq d+1$ is sufficient and necessary. For $\H=\C$, it is known that $N=2d$ generic measurements are sufficient for almost phase retrieval (see \cite{BCE06}). However, one still does not know whether $N=2d$ is tight or not.

\subsection{Our Contribution} In this paper we establish a general framework for the almost everywhere matrix recovery problem. Under our framework they are all unified under matrix recovery. One representative result in the paper is the following theorem on almost everywhere rank $r$ matrix recovery:

\begin{theo}   \label{theo-matrix-rand}
      Assume that $1 \leq r \leq \min(p,q)$ and $N>(p+q)r-r^2$. Let $\A=(A_j) \in (\H^{p\times q})^N$ be randomly chosen under an absolutely continuous probability distribution in $(\H^{p\times q})^N$, where $\H=\C$ or $\R$. Then with probability one $\A$  has  the almost everywhere rank $r$ matrix recovery property in $\H^{p\times q}$.
\end{theo}

In Theorem \ref{theo-matrix-rand} there is no constraint on the measurement matrices. However, often restrictions are put on these measurements. This turns out not to be an obstacle in general. Theorem \ref{theo-matrix-rand} is actually a special case of the following general theorem:

\vspace{2mm}
\begin{theo}   \label{theo-GenericMatrix}
       Let $V_j \subseteq \H^{p\times q}$ be algebraic varieties and $A_j\in V_j$ for $1 \leq j \leq N$, where $\H=\C$ or $\R$. Set $\A=(A_j)_{j=1}^N$. If $N<(p+q)r-r^2$ then $\A$ does not have the almost everywhere rank $r$ matrix recovery property in $\H^{p\times q}$. If $N > (p+q)r-r^2$  then a generic $\A$ in $V_1 \times \cdots \times V_N$  has  the almost everywhere rank $r$ matrix recovery property in $\H^{p\times q}$ if:
\begin{itemize}
  \item[\rm (A)]~~$V_j={\mathcal M}_{p\times q, r_j}(\H)$ and $1 \leq r_j \leq \min(p,q)$ for $1 \leq j \leq N$.
  \item[\rm (B)]~~For $\H=\R$ and $q =p=d$, $V_j$ is the set of $d\times d$ orthogonal matrices.
  \item[\rm (C)]~~For $\H=\R$ and $q =p=d$, $V_j$ is the set of $d\times d$ orthogonal projection matrix of rank $r_j\in [1,d-1]$.
\end{itemize}
\end{theo}

The theorem holds for far more broad classes of sets $V_j$'s. Later we shall show ways to establish this type of results in general, including some of the basic algebraic geometry tools needed for the study.  Theorem \ref{theo-GenericMatrix}  considers the case where $N\neq (p+q)r-r^2$. For the case where $N=(p+q)r-r^2$, we pose the following conjecture:

\vspace{0.5cm}
{\bf Conjecture.}
{\em  Suppose that $N=(p+q)r-r^2$ and $\H=\C$ or $\R$. The following statements hold:

(a) There exists $\A=(A_j)_{j=1}^N\subset \H^{p\times q}$  which has  the almost everywhere rank $r$ matrix recovery property in $\H^{p\times q}$.

(b) Set ${\mathcal P}:=\{\A=(A)_{j=1}^N\in \H^{N(p\times q)}: \A  \text{ has the almost everywhere rank $r$ matrix}\\ \text{ recovery property in } \H^{p\times q}  \}$. Then ${\mathcal P}$ is not dense in $\H^{N(p\times q)}$.
}
\vspace{0.5cm}

The paper is organized as follows. In Section 2, after introducing some results and notations from
elementary algebraic geometry, we present Theorem \ref{theo-admissible} which is often used in this study.
In Section 3, we prove that $\A=(A_j)_{j=1}^N $
with $A_j\in V_j\subset \H^{p\times q}$ has the almost everywhere $\M$-recovery property
if $N>\dim (\M)$ and $V_j$ satisfies  the admissible condition (see Definition \ref{defi-admissible}). We furthermore show that $\A$ does not have the almost everywhere $\M$-recovery property if $N<\dim(\M)$.
In Section 4, we prove the algebraic varieties introduced in Theorem \ref{theo-GenericMatrix}
satisfy the admissible condition. This implies Theorem \ref{theo-matrix-rand} and
 Theorem \ref{theo-GenericMatrix}.  We furthermore use the results in Section 3 to study the minimal
 measurement number for the recovery of Hermitian low rank matrices from rank one measurements.

%

\section{ Background from algebraic geometry}
\setcounter{equation}{0}

There is a strong connection between rank $r$ matrix recovery and the classical dimension theory in algebraic geometry. Such connection has been employed to study both matrix recovery and phase retrieval (see  \cite{Xu15, WaXu16, CEHV15}). Not surprisingly, this connection also plays a key role  for almost everywhere matrix recovery.  Before proceeding to the main results, we first introduce some basic notations related to projective spaces and varieties.

 For any complex vector space $X$ we shall use $\PP(X)$ to denote the induced projective space, i.e. the set of all one dimensional subspaces in $X$. As usual for each $\vx\in X$ we use $[\vx]$ to denote the induced elements in $\PP(X)$. Similarly, for any subset $S\subset X$ we use $[S]$ or $\PP(S)$ to denotes its projectivization in $\PP(X)$. Throughout this paper, we say $V\subset \C^d$ is a projective variety if $V$ is the locus of a collection of homogeneous  polynomials in $\C[\vx]$.
Strictly speaking a projective variety lies in $\PP(\C^d)$ and is the projectivization of the zero locus of a collection of homogeneous polynomials. But like in \cite{WaXu16}, when there is no confusion the phrase {\em projective variety in $\C^d$} means an algebraic variety in $\C^d$ defined by homogeneous polynomials. We shall use {\em projective variety in $\PP(\C^d)$} to describe a true projective variety. Note that sometimes it is useful to consider the more general quasi-projective varieties. A set $U \subset \C^d$ is a {\em quasi-projective variety} if there exist two projective varieties $V$ and $Y$ with $Y\subset V$ such that $U=V\setminus Y$. The concept of dimension for a quasi-projective variety in $\C^d$ is very well defined, and can be found in any standard algebraic geometry text such as \cite{alge}.

In studying almost everywhere  $\M$-recovery, we shall focus entirely on those $\M$ that are algebraic varieties in $\H^{p\times q}$. Note that the set $\M_{p\times q,r}(\H)$ is a projective variety as ${\rank}(Q)\leq r$ is  equivalent to the vanishing of all $(r+1)\times (r+1)$ minors of $Q$.  It is called a {\em determinantal variety} and has $\dim_\H({\mathcal M}_{p\times q,r}(\H))=(p+q)r-r^2$ \cite[Prop. 12.2]{alge}.

In \cite{WaXu16} the notion of an admissible algebraic variety with respect to a family of linear functions was introduced. The concept is equally useful in this paper.

\begin{defi}[\cite{WaXu16}]   \label{defi-admissible}  {\rm
   Let $V$ be the zero locus of a finite collection of homogeneous polynomials in $\C^{d}$ with $\dim (V)>0$ and let $\{\ell_\alpha(\vx): \alpha\in I\}$ be a family of (homogeneous) linear functions. We say $V$ is {\em admissible} with respect to $\{\ell_\alpha(\vx): \alpha \in I\}$ if  $\dim (V\cap \{\vx\in \C^d:\ell_\alpha(\vx)=0\}) <\dim (V)$ for all $\alpha\in I$.
}
\end{defi}

It is well known in algebraic geometry that if $V$ is irreducible in $\C^d$ then $\dim (V\cap Y) = \dim(V)-1$ for any hyperplane $Y$ that does not contain $V$. Thus the above admissible condition is equivalent to the property that no irreducible component of $V$ of dimension $\dim (V)$ is contained in any hyperplane $\ell_\alpha(\vx)=0$. In general without the irreducibility condition, admissibility is equivalent to that for a generic point $\vx\in V$, any small neighborhood $U$ of $\vx$ has the property that $U\cap V$ is not completely contained in any hyperplane $\ell_\alpha(\vx)=0$.  The following theorem extends a result in \cite{WaXu16}, and will play a key role in our paper.

\begin{theo}  \label{theo-admissible}
For $j=1, \dots, N$ let $L_j:\C^{n} \times \C^{m}\rightarrow \C$ be bilinear functions  and $V_j$ be projective varieties in $\C^n$.
Set $V := V_1\times \dots \times V_N \subseteq (\C^{n})^N$. Let $W, Y\subset \C^{m}$ be a projective variety in $\C^m$, $W \setminus Y$ be a quasi-projective variety.
For each fixed $j$, assume that $V_j$ is admissible with respect to the linear functions $\{f^\vw (\cdot)=L_j(\cdot,\vw):~\vw\in W\setminus Y\}$.
\begin{itemize}
\item[(A)]~~Assume that $N \geq \dim (W)$. There exists an algebraic subvariety $Z \subseteq V$ with $\dim (Z) < \dim (V)$ such that for any $\vx =(\vv_j)_{j=1}^N \in V\setminus Z$, the subvariety  $W_{\vx}$ given by
$$
    W_{\vx} := \Bigl\{\vw\in W \setminus Y:~L_j(\vv_j,\vw) = 0 \mbox{~for all $1\leq j \leq N$}\Bigr\}
$$
is the empty set.
\item[(B)]~~Assume that $N < \dim (W)$. There exists an algebraic subvariety $Z\subset V$  with $\dim (Z) < \dim (V)$ such that for any $\vx =(\vv_j)_{j=1}^N \in V\setminus Z$,  the subvariety  $W_{\vx}$ given by
$$
    W_{\vx} := \Bigl\{\vw\in W \setminus Y:~L_j(\vv_j,\vw) = 0 \mbox{~for all $1\leq j \leq N$}\Bigr\}
$$
has $\dim (W_{\vx}) = \dim (W)-N$.
\end{itemize}
\end{theo}

\Proof   We first prove (A). For $\vx=(\vv_j)_{j=1}^N\in V$, define $\Phi_{\vx} : W \setminus Y \to \C^N$ by $\Phi_{\vx}(\vw) = (L_j(\vv_j, \vw))_{j=1}^N$. Let $\G$ be the subset of $[V] \times [W \setminus Y] \subset \PP ((\C^n)^N) \times \PP (\C^m)$ such that $([\vx], [\vw]) \in \G$ if and only if $\Phi_{\vx} (\vw) = 0$, i.e. $L_j (\vv_j, \vw) = 0$ for all $j$.
We can view $\G$ as a quasi-projective variety via Segre embedding \cite[Page 27]{alge}.
 Note that $\G$ is a projective variety of $\PP((\C^n)^N) \times \PP(\C^m)$. We consider its dimension. Let $\pi_1$ and $\pi_2$ be projections from $\PP ((\C^n)^N) \times \PP(\C^m)$ onto the first and the second coordinates, respectively, namely
$$
\pi_1([\vx], [\vw]) = [\vv_1, \ldots, \vv_N],\qquad \pi_2([\vx], [\vw]) = [\vw].
$$
It is easy to check that $\pi_2 (\G) = [W \setminus Y]$, the projection of $W \setminus Y$. Thus $\dim (\pi_2 (\G)) = \dim (W \setminus Y) - 1$.

We next consider the dimension of the preimage of the $\pi_2^{-1} ([\vw_0]) \subset \PP((\C^n)^N)$ for a fixed $[\vw_0] \in \PP(\C^m)$. Let $V'_j := V_j \cap H_j$ where $H_j : = \{ \vy \in \C^n : L_j(\vy, \vw_0)=0\}$ is a hyperplane. The admissibility property of $V_j$ implies that $\dim(V_j')=\dim (V_j) - 1$. Hence after projectivization the preimage $\pi_2^{-1}([\vw_0])$ has dimension
$$
\dim (\pi_2^{-1}([\vw_0])) = \sum_{j=1}^N (\dim (V_j) - 1) - 1 = \dim(V) - N - 1.
$$
According to Cor.11.13 in \cite{alge}, we have
\begin{eqnarray*}
      \dim (\G) &=& \dim (\pi_2(\G)) + \dim(\pi^{-1}([\vw_0])) \\
      &=& (\dim (W \setminus Y) -1) + (\dim (V) -N -1)\\
      &=& \dim (V) + \dim (W \setminus Y) -N -2\\
      &\leq& \dim (V) + \dim (W) -N -2.
\end{eqnarray*}
If $N \geq \dim (W)$ then
$$
\dim (\pi_1(\G)) \leq \dim (\G) = \dim (V) + \dim (W) -N -2 \leq \dim (V) -2.
$$
Note that $\pi_1(\G)$ is itself a projective variety. Let $Z$ be the lift of $\pi_1(\G)$ into the vector space $(\C^n)^N$. Then
$$
\dim (Z) \leq \dim (V)-1.
$$
The definition of $Z$ implies that $W_\vx$ is an empty set provided $\vx\in V\setminus Z$.

Next we prove (B). Let $K = \dim (W \setminus Y)$. Noting $K>N$, we augment $\{V_j\}_{j=1}^N$ and $\{L_j(\vv,\vw)\}_{j=1}^N$ to $\{V_j\}_{j=1}^K$ and $\{L_j(\vv,\vw)\}_{j=1}^K$ via $V_j =V_1$ and $L_j(\vv,\vw)=L_1(\vv,\vw)$ for all $j>N$. Set $\hat V = V_1\times \dots \times V_K \subseteq (\C^{n})^K$. By (1) there exists a subvariety $\hat Z$ of $\hat V$ with $\dim (\hat Z) <\dim (\hat V)$ such that for any $\hat\vx=(\vv_j)_{j=1}^K \in \hat V\setminus \hat Z$ and $\vw\in W \setminus Y$, we have $L_j(\vv,\vw) \neq 0$ for some $j\in [1,K]$. Now consider the sequence of nested varieties with $W_{\hat\vx,0} = W$ and
$$
    W_{\hat\vx, k} := \Bigl\{\vw \in W \setminus Y:~L_j(\vv_j,\vw)=0 \mbox{~for all $1\leq j \leq k$}\Bigr\}, \quad k=1,\ldots,K.
$$
Thus the above is equivalent to $W_{\hat\vx,K} = \emptyset$ provided  $\hat\vx \in \hat V\setminus \hat Z$.

Since for each fixed $\vv_j$ the equation $L_j(\vv_j, \vw)=0$ defines a hyperplane $H$ in $\C^{m}$, it is well known that $\dim(U\cap H) \geq \dim(U)-1$ for any variety $U$ in $\C^{m}$. Then we have a decreasing sequence of subvarieties of $\C^{m}$
$$
    W \setminus Y = W_{\hat\vx, 0} \supseteq W_{\hat\vx, 1} \supseteq W_{\hat\vx, 2} \supseteq \cdots  \supseteq W_{\hat\vx, K} =\emptyset.
$$
Now $\dim(W_{\hat\vx,0})=K$. By Krull's Principal Ideal Theorem, at each step the dimension can only be reduced by at most 1, we must thus have $\dim(W_{\hat\vx, k-1})-1 = \dim(W_{\hat\A,k})$ for $1 \leq k \leq K$. It follows that  $\dim(W_{\hat\vx,N})= \dim (W)-N=K-N$.

Thus for any $\vx=(\vv_j)_{j=1}^N\in V$, if there exists $\vv_j \in V_j$ for $N<j\leq K$ such that $\hat\vx = (\vv_j)_{j=1}^K\in \hat V\setminus\hat Z$  we must have $\dim(W_{\hat\vx,N})= K-N$. Since $W_{\hat\vx,N} = W_\vx$ we then have $\dim(W_{\vx})= K-N$. Finally, let $Z=\{\vx=(\vv_j)_{j=1}^N\}\subset V$ be those such that there exists no such extensions $\hat\vx\in\hat V\setminus\hat Z$. We have
$$
     Z=\Bigl\{\vx=(\vv_j)_{j=1}^N\in V:~\hat\vx=(\vv_j)_{j=1}^K\in\hat Z ~\mbox{for any}~\vv_j \in V_j, j>N\Bigr\}.
$$
Since $\hat Z$ is variety in $(\C^n)^K$, $Z$ is a variety. Clearly it has $\dim(Z) <\dim (V)$, for otherwise we would have $\dim(\hat Z) = \dim(\hat V)$, which is a contradiction.
\eproof
\begin{remark}
The approach we take for proving (A) in Theorem \ref{theo-admissible} is similar to one taken in \cite{Bilinear} to study injectivity conditions for bilinear inverse problems. However, part (A) is not enough for the aim of this paper, and hence we develop new approach to prove part (B).
\end{remark}

For real matrix recovery we  need to consider real projective varieties.  Here we introduce some notations. Let $V$ be a variety in $\C^d$.  We shall use $V\cap \R^d$ to denote the real points of $V$. As a real variety we can define the {\em real dimension} of $V\cap \R^d$, see \cite{alge} and \cite{real}.  A key fact is that for a variety $V$ we have $\dim_\R(V\cap \R^d)\leq \dim(V)$ (see Section 2.1.3 in \cite{E15} and \cite{WaXu16}). This also holds for a quasi-projective variety since the proof uses only local properties of $V$ (see \cite{WaXu16}).

A particularly important class of projective varieties for our study are those $V\subseteq\C^d$ such that $\dim_\R(V\cap\R^d)=\dim (V)$. For example, $V={\mathcal M}_{p\times q,r}(\C)$ in $\C^{p\times q}$ has this property. This class is especially useful for real matrix recovery.

\section{Almost Everywhere Matrix Recovery: General Results}
\setcounter{equation}{0}

In this section we consider the problem of almost everywhere matrix recovery. At the same time,  we also prove results on the classical matrix recovery (i.e. everywhere matrix recovery). Let $\M$ be a projective variety in $\H^{p\times q}$ such as $\M = {\mathcal M}_{p\times q,r}(\H)$, and let $P\in  {\mathcal M}$. We ask how many linear measurements are needed to recover $P$ for all $P\in\M$, and how many linear measurements  are needed to recover $P$ for {\em almost} all $P\in\M$.

\begin{theo} \label{theo-aeMatrixRecovery}
Assume that $\A=(A_j)_{j=1}^N\in (\H^{p\times q})^N$ where $\H =\C$ or $\R$. Let $\M$ be a projective variety in $\C^{p\times q}$ with $\dim(\M)=K$ and $Y_\A $ be
$$
    Y_\A := \Bigl\{(P,Q):~P,Q\in {\mathcal M},~ P\neq Q, ~\Tr(A_j^T(P-Q))=0 \mbox{~for $1\leq j \leq N$}\Bigr\}\subset \H^{p\times q}\times \H^{p\times q}.
$$
\begin{itemize}
\item[\rm (A)]~~For $\H=\C$, if the (complex) quasi-projective  variety $Y_\A$  has $\dim(Y_\A) <K$ then $\A$ has the almost everywhere $\M$-recovery property.
\item[\rm (B)]~~For $\H=\R$ let $\M_\R =\M\cap\R^{p\times q}$. If $\dim_\R(\M_\R)=K$ and $\dim_\R(Y_\A) <K$ then $\A$ has the almost everywhere $\M_\R$-recovery property.
\end{itemize}
\end{theo}
\Proof   First we consider the case $\H=\C$. Let $Z$ denote the set of matrices $P\in {\mathcal M}$ such that there exists a $Q\neq P$ in ${\mathcal M}$ such that $\Tr(A_j^T P)=\Tr(A_j^T Q)$ for all $1 \leq j \leq N$. The goal is to show that $Z$ is a null set in ${\mathcal M}$. Observe that the set $Z$ is the projection of $Y_{\A}$ onto the first coordinate. Since projections cannot increase dimension (see \cite{alge}[Cor.11.13]), it follows that $\dim (Z) < K$. Hence $Z$ is a null set in ${\mathcal M}$.

Now for $\H=\R$, we already stated that the real dimension of $Y_{\A}\cap \R^{p\times q}\times \R^{p\times q}$ is no larger than the (complex) dimension of $Y_{\A}$. Thus $\dim_\R(Y_\A) <K$. Let $Z_\R$ denote the set of matrices $P\in {\mathcal M}_\R$ such that there exists a $Q\neq P$ in ${\mathcal M}_\R$ with $\Tr(A_j^T P)=\Tr(A_j^T Q)$ for all $1 \leq j \leq N$. Again $Z_\R$ is the projection of $Y_{\A}\cap \R^{p\times q}$ onto the first coordinate. Since projections cannot increase dimension, it follows that $\dim_\R (Z_\R) < K$. Hence $Z_\R$ is a null set in ${\mathcal M}_\R$. The theorem is proved.
\eproof

Intuitively speaking, the larger the number of measurements is the smaller $\dim(Y_\A)$ will be. So the question is how many measurements do we need to reach $\dim(Y_\A) <K$. Our next theorem provides the answer for generic measurements restricted to projective varieties. First we establish a very intuitive lemma.

\begin{lem}  \label{lem:noninjective}
   Let $\Phi: U \lra \R^n$ be a $C^1$ map, where $U\subseteq \R^m$ is an open set and $m>n$. Then $\Phi$ cannot be almost everywhere injective.
\end{lem}
\Proof  Let $J :=(\partial \phi_i/\partial x_j)$ be the Jacobian matrix of $\Phi=(\phi_1, \dots,\phi_n)^T$. Let $r$ be the maximal rank of $J$ on $U$. The set of points in $U$ at which the rank of $J$ is $r$ is an open set in $U$, and we shall show that $\Phi$ is not almost everywhere injective on this set. So without loss of generality we may assume that $\rank(J)=r$ everywhere on $U$.

We first consider the case where $r=n$.  For any $\vx_0\in U$ let $\vy_0=\Phi(\vx_0)$. Without loss of generality again we may assume that the first $n$ columns of $J(\vx_0)$ are linearly independent. Set $F(\vx)=\Phi(\vx)-\vy_0$. By the Implicit Function Theorem there exist functions $\psi_{n+1}(x_1, \dots, x_n), \dots, \psi_m(x_1, \dots, x_n)$ in a small neighborhood of $\vx_0$ such that
\[
   F(x_1, \dots, x_n, \psi_{n+1}(x_1, \dots, x_n), \dots, \psi_m(x_1, \dots, x_n))=0,
\]
namely
\[
   \Phi(x_1, \dots, x_n, \psi_{n+1}(x_1, \dots, x_n), \dots, \psi_m(x_1, \dots, x_n))=\vy_0.
\]
Thus $\Phi^{-1}(\Phi(\vx_0))$ contain more than just $\vx_0$. It follows that $\Phi$ is not almost everywhere injective.

We next consider the case $r<n$. By the Rank Theorem (see \cite{KrPa-book}, Theorem 3.5.1), for any $\vx_0\in U$ there is a decomposition $\R^n = V\oplus \tilde V$ where $V, \tilde V$ are linear subspaces of $\R^n$ with $\dim (V)=r$ and $\dim (\tilde V)= n-r$, such that we can write $\Phi(\vx)$ in a small neighborhood $W_{\vx_0}$ of $\vx_0$ as
$$
      \Phi(\vx) = \Phi_1(\vx)+\Phi_2(\vx),  \mhsp \Phi_1(\vx) \in V,\, \Phi_2(\vx) \in\tilde V,
$$
with the property that the value of $\Phi_2(\vx)$ is uniquely determined by the value of $\Phi_1(\vx)$. In other words, if $\Phi_1(\vx) = \Phi_1(\vx')$ then $\Phi_2(\vx) = \Phi_2(\vx')$ for any $\vx,\vx'\in W_{\vx_0}$. It follows that $\Phi$ is almost everywhere injective on $W_{\vx_0}$ if and only if $\Phi_1$ is almost everywhere injective on $W_{\vx_0}$. If the Jacobian of $\Phi_1$ has rank $r$ we have already shown from the first case that $\Phi_1$ cannot be almost everywhere injective, and hence nor can $\Phi$. But if the Jacobian of $\Phi_1$ has rank $<r$ then we can repeat the argument, and eventually yields that $\Phi$ cannot be almost everywhere injective.
\eproof

Through out the rest of this paper, we set
\[
\Delta \M\,\,:=\,\, \{ \vx-\vy : \text{ for all } \vx, \vy \in \M\}.
\]
\begin{theo}  \label{matrix-complex-generic}
    Let $\M$  and $V_j$ be projective varieties in $\C^{p\times q}$, $j=1, \dots, N$. Assume that  each $V_j$ is admissible with respect to the maps $\{ L_P:~P\in \Delta \M, P\neq 0\}$ where $L_P(Q):=\Tr(P^T Q)$. Then for $\A=(A_j)_{j=1}^N$ where $A_j\in V_j$ we have
    \begin{itemize}
      \item[\rm (A)]~~If $N<\dim(\M)$ then $\A$ does not have the almost everywhere $\M$-recovery property. On the other hand, if $N > \dim(\M)$ then a generic $\A=(A_j)_{j=1}^N$ in $V_1\times V_2 \times \cdots\times V_N$ has the  almost everywhere $\M$-recovery property.
      \item[\rm (B)]~~ Suppose that $\Delta\M$ is a projective variety. If $N<\dim(\Delta \M)$ then $\A$ does not have the $\M$-recovery property. On the other hand, if $N \geq \dim(\Delta \M)$ then a generic $\A=(A_j)_{j=1}^N$ in $V_1\times V_2 \times \cdots\times V_N$ has the   $\M$-recovery property.
    \end{itemize}
\end{theo}
\Proof  Let $K=\dim(\M)$. First we prove (A). If $N <K$ then the map $L_\A$ given in (\ref{eq:trace-map}) maps smoothly the higher dimensional manifold $\M$ to the lower dimensional one $\C^N$. For the aim of contradiction, we suppose  that  $L_\A$ is almost everywhere injective. By looking at $\M$ locally we see that there exists a smooth map $\Phi$ from a ball $B$ in $\C^K\cong \R^{2K}$ to $\C^N\cong \R^{2N}$ that is almost everywhere injective. But this is impossible by Lemma \ref{lem:noninjective}.

Now for $N >K$ let $X \subset \C^{p\times q}\times \C^{p\times q}$ be the quasi-projective variety
$$
    X := \Bigl\{(P,Q):~P,Q\in \M,~ P\neq Q\Bigr\}.
$$
For each $(P,Q) \in X$ denote $\psi_{(P,Q)}(A)=\Tr(A^T(P-Q))$. As in Theorem \ref{theo-aeMatrixRecovery} set
$$
    Y_\A := \Bigl\{(P,Q):~P,Q\in \M,~ P\neq Q, \Tr(A_j^T(P-Q))=0
                                 \mbox{~for $1\leq j \leq N$}\Bigr\}.
$$
Since each $V_j$ is admissible with respect to the maps $\{\psi_{(P,Q)}:(P,Q)\in X\}$. By Theorem \ref{theo-admissible} we have $\dim(Y_\A)=\dim(X)-N<2K-K =\dim(\M)$. It follows from Theorem \ref{theo-aeMatrixRecovery} that $\A$ has the  almost everywhere $\M$-recovery property.

For (B) it is essentially proved in \cite{WaXu16}. We quickly recap it here. For $N<\dim(\Delta \M)$ the dimension of the projective variety $U:=\bigl\{Q \in \Delta \M:~\Tr(A_j^T Q)=0, j=1,\ldots,N \bigr\}$ is no less than $\dim(\Delta \M)-N>0$. This is because in the complex projective space, through intersection with a hyperplane such as the one given by $\Tr(A_j^T Q)=0$, the  dimension of any projective variety can be reduced by at most one. Thus there exists a $Q = Q_1 -Q_2 \neq 0$ with $Q_1,Q_2\in\M$ such that $\Tr(A_j^T Q_1)=\Tr(A_j^T Q_2)$ for all $j$. Hence $\A$ does not have the $\M$-recovery property.

In the case $N \geq \dim(\Delta \M)$, we apply Theorem \ref{theo-admissible} with $W=(\Delta \M)\setminus \{0\}$ and $L_j(A, Q):=\Tr(A^T Q)$. Let $V= V_1\times V_2 \times \cdots\times V_N$. Then there exists a variety $Z \subset V$ with $\dim(Z)<\dim(V)$ such that for all $\A\in V\setminus Z$ there exists no $Q\in W$ with the property $\Tr(A_j^T Q)=0$ for all $j$. Thus a generic $\A \in V$ has the   $\M$-recovery property.
\eproof

For the real case the above theorem can be extended. First, for any real variety $V \subseteq \R^d$ it has a natural extension to a variety in $\C^d$. The ideal $I_\R(V)$ defining $V$ generates an ideal $I_\C(V)$ in $\C^d$, and the variety corresponding to $I_\C(V)$ will be our extension, which we denote it by $\bar V$. Note that $V$ is clearly the restriction of $\bar V$ to $\R^d$, namely $V = \bar V_\R$ using the terminology in this paper.
\begin{theo}  \label{matrix-real-generic}
    Let $\M$  and $V_j$ be projective varieties in $\R^{p\times q}$, $j=1, \dots, N$. Let each $\bar V_j$ be admissible with respect to the maps $\{ L_P:~P\in \Delta \bar\M, P\neq 0\}$ where $L_P(Q):=\Tr(P^T Q)$. Assume further that $\dim_\R(\M)=\dim(\bar\M)$ and $\dim_\R(V_j)=\dim(\bar V_j)$ for all $j$. Then for $\A=(A_j)_{j=1}^N$ where $A_j\in V_j$ we have
    \begin{itemize}
      \item[\rm (A)]~~If $N<\dim_\R(\M)$ then $\A$ does not have the almost everywhere $\M$-recovery property. On the other hand, if $N > \dim_\R(\M)$ then a generic $\A=(A_j)_{j=1}^N$ in $V_1\times V_2 \times \cdots\times V_N$ has the  almost everywhere $\M$-recovery property.
      \item[\rm (B)]~~Assume additionally that $\dim_\R(\Delta \M)=\dim(\Delta \bar\M)=L$. If $N \geq L$ then a generic $\A=(A_j)_{j=1}^N$ in $V_1\times V_2 \times \cdots\times V_N$ has the   $\M$-recovery property.
    \end{itemize}
\end{theo}
\Proof  Denote $K:=\dim(\M)$ and $V:=V_1\times V_2 \times \cdots\times V_N$. First we prove (A). If $N <K$ then the map $L_\A$ given in (\ref{eq:trace-map}) maps smoothly the higher dimensional manifold $\M$ to the lower dimensional one $\R^N$. Again if $L_\A$ is almost everywhere injective, by looking at $\M$ locally we see that there exists a smooth map $\Phi$ from a ball $B$ in $\R^K$ to $\R^N$ that is almost everywhere injective. This is a contradiction.

Now for $N >K$ we lift $\M$ and $V$ into the complex projective space to $\bar\M$ and $\bar V$. Let $X \subset \C^{p\times q}\times \C^{p\times q}$ be the quasi-projective variety
$$
    X := \Bigl\{(P,Q):~P,Q\in \bar\M,~ P\neq Q\Bigr\}.
$$
For each $(P,Q) \in X$ denote $\psi_{(P,Q)}(A)=\Tr(A^T (P-Q))$. As in Theorem \ref{theo-aeMatrixRecovery} set
$$
    \bar Y_\A := \Bigl\{(P,Q):~P,Q\in \bar\M,~ P\neq Q, \Tr(A_j^T(P-Q))=0
                                 \mbox{~for $1\leq j \leq N$}\Bigr\},
$$
and let $Y_\A = \bar Y_\A \cap \R^{p\times q}\times \R^{p\times q}$ be its restriction to the reals.
Note that each $\bar V_j$ is admissible with respect to the maps $\{\psi_{(P,Q)}:(P,Q)\in X\}$. By Theorem \ref{theo-admissible} there exists a subvariety $\bar Z\subset \bar V$ with $\dim (\bar Z)<\dim (\bar V)$ such that for any $\A \in \bar V\setminus \bar Z$ we have $\dim(\bar Y_\A)=\dim(X)-N<2K-K =\dim(\bar\M)$. By assumption we have $\dim_\R (V) = \dim (\bar V)$ so the restriction $Z={\bar Z}_\R$ of $\bar Z$ to the reals must have $\dim_\R (Z)<\dim (V)$. Furthermore,  $\dim_\R(Y_\A) \leq \dim(\bar Y_\A)<K$. It follows from Theorem \ref{theo-aeMatrixRecovery} that any $\A\in V\setminus Z$ has the  almost everywhere $\M$-recovery property. This proves (A).

For (B) we follow the same strategy, which has been used for phase retrieval in \cite{WaXu16}. We apply Theorem \ref{theo-admissible} with $\bar W=(\Delta \bar\M)\setminus \{0\}$ and $L_j(A, Q):=\Tr(A^T Q)$. Let $V, \bar V$ be as in part (A). Since $N \geq \dim(\Delta \bar\M)$ it follows from Theorem \ref{theo-admissible} that there exists a variety $\bar Z \subset \bar V$ with $\dim (\bar Z)<\dim (\bar V)$ such that for any $\A \in \bar V\setminus \bar Z$ there exists no $Q\in \bar W$ with the property $\Tr(A_j^T Q)=0$ for all $j$. Now
$$
    \dim_\R(Z) \leq \dim(\bar Z) < \dim (\bar V) = \dim_\R(V).
$$
Thus any $\A \in V\setminus Z$ has the   $\M$-recovery property, which means a generic $\A\in V$ has the $\M$-recovery property.
\eproof

\section{Cases of Almost Everywhere Matrix Recovery}
\setcounter{equation}{0}

In this section we provide several applications of almost everywhere matrix recovery on algebraic varieties of interest.
\subsection{Algebraic varieties satisfying the admissibility condition }
According to Theorem \ref{matrix-complex-generic}, the admissibility condition plays a key role in studying the almost everywhere matrix recovery. Then, we list many algebraic varieties as follows which satisfy this condition:

\begin{prop}  \label{prop:adm}
     Let $V$ be one of the following projective varieties in $\C^{p\times q}$. Then $V$  is admissible with respect to any set of nontrivial linear functions on $\C^{p\times q}$:
\begin{itemize}
\item[\rm (A)]~~$V =  {\mathcal M}_{p\times q,s}(\C)$, the set of all $p\times q$ complex matrices of rank $s$ or less, where $1 \leq s \leq \min(p,q)$.
\item[\rm (B)]~~$q \geq p$ and $V$ is the set of all scalar multiples of matrices $P$ satisfying $PP^T=I$.
\item[\rm (C)]~~$p=q$ and $V$ is the set of  all scalar multiples of projection matrices $P$, i.e.  $P^2=P$.
\end{itemize}
\end{prop}
\Proof   (A) To this end, we just need show that for a generic $P_0\in V$ and any nontrivial linear function $f$ on $\C^{p\times q}$ we have $f(P) \not\equiv 0$ on any neighborhood of $P_0$. Note that there exists a nonzero $Q_0\in\C^{p \times q}$ such that $f(P)=\Tr(Q_0^TP )$ for all $P$. If $ \Tr(Q_0^TP_0 )\neq 0$ then we are done. For the case $ \Tr(Q_0^TP_0 )=0$, there always exist two matrices $S_1\in \C^{q\times q}, S_2\in \C^{p\times p}$ so that $ \Tr(Q_0^TS_2P_0S_1)\neq 0$. Take $P_{t}=(I+tS_2)P_0(I+tS_1)\in V$. Then
$$
     \Tr(Q_0^TP_{t} ) = t^2 \Tr(S_2 P_0 S_1) + t \Tr(S_2 P_0)
                                             + t \Tr(P_0 S_1).
$$
Clearly $  \Tr(Q_0^TP_{t}) \not \equiv 0$ for sufficiently small $t\neq 0$. We thus arrive at the conclusion. \\
(B) It is sufficient to show that a generic point $ P_0 \in V $ and any nonzero $ Q_0 \in \C^{p \times q} $ we must have $ \Tr(Q_0^TP) \not\equiv 0 $ in any small neighborhood of $ P_0 $ in $ V $.
If $ Q_0^TP_0  \neq 0 $, then set $ P_t := P_0 e^{tS}  $, where $ S $ is a skew-symmetric $ q \times q$ matrix.
 A simple observation is that $ P_t \in V$.
 Then all we need to show is that for some $ S $ and arbitrarily small $ t \neq 0 $, we have $ \Tr(Q_0^TP_t) \not \equiv 0 $.
Then
$$
\Tr(Q_0^TP_t ) = \Tr(Q_0^T P_0e^{tS}) = \Tr(Q_0^TP_0 ) + \sum_{n=1}^{+\infty} \frac{1}{n!} t^n \Tr( Q_0^TP_0 S^n).
$$
If $ \Tr(Q_0^TP_0 ) \neq 0 $ then we are done.  We next assume that $ \Tr(Q_0^TP_0 ) = 0 $. To this end, we show there is a $ S_0 $ such that $ \Tr(Q_0^T P_0 S_0) \neq 0 $. \\
We first consider the case where  $ Q_0^TP_0  $ is not a symmetric matrix. Then there exists $ 1 \leq i < j \leq q $ such that $ (Q_0^TP_0 )_{ij} \neq (Q_0^TP_0 )_{ji} $ and we can define $ S $ by setting all the entries to be zero except $ (S_0)_{ij} = 1 $ and $ (S_0)_{ji} = -1 $. Then we have $ \Tr(Q_0^T P_0S_0 ) = (Q_0^TP_0 )_{ji} - (Q_0^TP_0)_{ij} \neq 0 $. \\
If $ Q_0^TP_0  $ is a symmetric matrix, we claim that there exists a skew-symmetric matrix $ S_1 $ such that $ Q_0^TP_0e^{t_1 S_1}  $ is not symmetric for $ t_1 \in (0, 1] $. Then we can take $ P_{t_0, t_1} = Q_0^TP_0e^{t_0 S_0 + t_1 S_1} $ and the above statement will hold. To verify the claim, notice that
$$
Q_0^TP_0e^{t_1 S_1}  = Q_0^TP_0 + \sum_{n=1}^{+\infty} \frac{1}{n!} t_1^n Q_0^T P_0S_1^n,
$$
and it is sufficient to show that there is a skew-symmetric matrix $ S_1 $ such that $ Q_0^TP_0S_1 $ is not symmetric. Since $ Q_0^TP_0  \neq 0 $, there exists $ 1 \leq i, j \leq q $ such that $ (Q_0^TP_0)_{ij} \neq 0 $, and then choose $ 1 \leq k \leq q $ such that $ k \neq i $, $ k \neq j $, and define $ S_1 $ by setting all the entries to be zeros except $ (S_1)_{jk} = 1 $ and $ (S_1)_{kj} = -1 $. Then we have
\[
 (Q_0^TP_0 S_1)_{ik} = (Q_0^TP_0 )_{ij} \neq 0 = (Q_0^T P_0 S_1)_{ki}.
\]
It remains to discuss the case where $ Q_0^TP_0  = 0 $. In that case, since $ P_0, Q_0 \neq 0 $, we claim that there exists a skew-symmetric matrix $ S_2\in \C^{p\times p} $  such that $ Q_0^T e^{t_2 S_2} P_0 \neq 0 $ for any $ t_2 \in (0, 1] $. Then we can set $ P_{t_0, t_1, t_2} =  Q_0^T e^{t_2 S_2} P_0 e^{t_0 S_0 + t_1 S_1}$ and the above statement will hold. To verify the claim, notice that
$$
Q_0^T e^{t_2 S_2} P_0 =  Q_0^TP_0 + \sum_{n=1}^{+\infty} \frac{1}{n!} t^n Q_0^T S_2^n P_0,
$$
and it is sufficient to show that there exists a skew symmetric matrix $ S_2 $ such that $ Q_0^T S_2 P_0 \neq 0 $. Assume the above claim does not hold. Since $ P_0, Q_0 \neq 0 $, we can choose $ k, l $ such that the $ k $-th row of $ Q_0^T $, denoted by $ (Q_0^T)^k $, and the $ l $-th column of $ P_0 $, denoted by $ (P_0)_l $, are nonzero. Then for any $ 1 \leq i < j \leq q $, if we define $ S_2 $ by setting all the entries to be zero except $ (S_2)_{ij} = 1 $ and $ (S_2)_{ji} = -1 $, and we will have $ (Q_0^T S_2 P_0)_{kl} =  (P_0)_{jl}(Q_0^T)_{ki} - (P_0)_{il}(Q_0^T)_{kj}  = 0 $. And we have
 \[
  ((Q_0^T)_{ki}, (Q_0^T)_{kj}) =\lambda_{ij,kl} ((P_0)_{il}, (P_0)_{jl})
 \]
 for any $ 1 \leq i < j \leq q $ such that $\abs{ (P_0)_{jl}}^2+ \abs{(P_0)_{il}}^2 \neq  0 $. Hence $ (Q_0^T)^k $ and $ (P_0)_l $ satisfy $ (Q_0^T)^k = \lambda_{kl} (P_0)_l^T $ where $  \lambda_{kl} \neq 0 $. Then we have $ (Q_0^TP_0 )_{kl} = \lambda_{kl} || (P_0)_l ||^2 \neq 0 $, which contradicts to the assumption that $ Q_0^TP_0  = 0 $. This completes the proof. \\
(C) Let $ d = p = q $. It is sufficient to show that at a generic point $ P_0 \in V $ and any nonzero $ Q_0 \in \C^{d \times d} $ we must have $ \Tr(Q_0^TP) \not\equiv 0 $ in any small neighborhood of $ P_0 $ in $ V $. \\
Since $ P_0^2 = P_0 $, there exists  a nonsingular matrix  $ R $ such that $ P_0 = R J_s R^{-1} $ where
\begin{align*}
    J_s =
    \begin{pmatrix}
        I_s & 0 \\
        0 & 0
    \end{pmatrix}
    \in \C^{d \times d}.
\end{align*}
The property of trace implies that $ \Tr(Q_0^TP_0 ) = \Tr(R^{-1}Q_0^TRJ_s) $. Hence, without loss of generality, we just need consider the case where $P_0=J_s$. Set $ P_t = (I+tS) J_s (I+tS)^{-1} $. Then all we need to show is that for some $ S $ and arbitrary small $ t \neq 0 $ we have $ \Tr(Q_0^TP_t) \neq 0 $. Since $ (I+tS)^{-1}  = \sum_{n=0}^{\infty} (-1)^n t^n S^n $, we have
$$
\Tr(Q_0^TP_t ) = \Tr(Q_0^TP_0) + \sum_{n=1}^{\infty} (-1)^{n-1} t^n \Tr(Q_0^T(SJ_s-J_sS)S^{n-1}).
$$
If there exists a $ S \in \C^{d \times d} $ such that $ \Tr(Q_0^T(SJ_s-J_sS)S^{n-1}) \neq 0 $ for some $ n \geq 1 $ then we are done. \\
For $ n = 1 $,
$$
\Tr(Q_0^T(SJ_s-J_sS)S^{n-1}) = \Tr(Q_0^T(SJ_s-J_sS)) = \Tr(S(J_sQ_0^T-Q_0^TJ_s)).
$$
We first consider the case where $ J_s Q_0^T - Q_0^T J_s \not\equiv 0 $. Then we can take $ S = (J_s Q_0^T - Q_0^T J_s)^* $ and obtain $ \Tr(S(J_s Q_0^T - Q_0^T J_s)) \neq 0 $. We are done. \\
We next only consider the case where $ J_s Q_0^T - Q_0^T J_s = 0 $. Then $ Q_0^T $ must have the form
\begin{align*}
    Q_0^T =
    \begin{pmatrix}
        Q_1 & 0 \\
        0 & Q_2
    \end{pmatrix}
\end{align*}
where $ Q_1 \in \C^{s \times s} $ and $ Q_2 \in \C^{(d-s) \times (d-s)} $. Consider now $ n=2 $ and we have
\begin{eqnarray*}
    Q_0^T(SJ_s - J_sS) S^{n-1}  &=&
    \begin{pmatrix}
        Q_1 & 0 \\
        0 & Q_2
    \end{pmatrix}
    \begin{pmatrix}
        0 & -S_{12} \\
        S_{21} & 0
    \end{pmatrix}
    \begin{pmatrix}
        S_{11} & S_{12} \\
        S_{21} & S_{22}
    \end{pmatrix}
   \\
    &=&
    \begin{pmatrix}
        -Q_1S_{12} S_{21}  & -Q_2S_{12} S_{22}  \\
        Q_1S_{21} S_{11}  & Q_2S_{21} S_{12}
    \end{pmatrix},
\end{eqnarray*}
which yields $ \Tr(Q_0^T(SJ_s-J_sS)S) = \Tr(-Q_1S_{12}S_{21}+Q_2S_{21}S_{12}) $. \\
Assume that $ Q_1, Q_2 $ are not both scalar multiples of identity matrices. Without loss of generality, suppose $ Q_1 \neq \lambda I_s $ for any $ \lambda \in \C $, where $ I_s \in \C^{s \times s} $ is the identity matrix. Then there exist $ \mathbf{u}, \mathbf{v} \in \C^s $ such that $ \mathbf{v}^*\mathbf{u} =0 $ but $ \mathbf{v}^* Q_1 \mathbf{u} \neq 0 $. Take $ S_{12} = \mathbf{u} \mathbf{x}^* $ and $ S_{21} = \mathbf{x} \mathbf{v}^* $ where $ \mathbf{x} \in \C^{d-s} $ and $ \mathbf{x} \neq 0 $. Then
$$
\Tr(Q_0^T(SJ_s-J_sS)S) = -\mathbf{v}^* Q_1 \mathbf{u} \|\mathbf{x}\|^2 \neq 0.
$$
We next consider the case where both $ Q_1, Q_2 $ are scalar multiples of identity matrices, that is, $ Q_1 = \lambda_1 I_s $ and $ Q_2 = \lambda_2 I_{d-s} $ where $ \lambda_1, \lambda_2 \in \C $. We claim that $ \lambda_1 \neq \lambda_2 $, otherwise we have $ Q_0 = \lambda I_d $ where $ \lambda \in \C $ and $ \lambda \neq 0 $. Then $ \Tr(Q_0^TP_0) = \lambda s  \neq 0 $, which contradicts to the assumption above. Thus we can take $ S_{12} = \mathbf{x} \mathbf{y}^* $ and $ S_{21} = \mathbf{y} \mathbf{x}^* $ where $ \mathbf{x} \in \C^s, \mathbf{x} \neq 0 $ and $ \mathbf{y} \in \C^{d-s}, \mathbf{y} \neq 0 $. Then
$$
\Tr(Q_0^T(SJ_s-J_sS)S) = (\lambda_2 - \lambda_1) \|\mathbf{x}\|^2 \|\mathbf{y}\|^2 \neq 0.
$$
This completes the proof.
\eproof

\begin{remark}
It should be noted that there are indeed some projective varieties $V$ that are not admissible with respect to certain class of linear functions on $\C^{p\times q}$. For example, if $V$ is the set of all symmetric $p\times p$ matrices, then $f(P)=\Tr(P^T Q) \equiv 0$ on $V$ for any skew-symmetric $Q$. Hence, the $V$ is {\em not} admissible with respect to $\{\Tr(\cdot Q):Q\in \C^{p\times p}, Q^T=-Q\}$. Nevertheless the admissibility condition for many application such as phase retrieval does hold, but often needs to be individually checked.
\end{remark}

The case most people study is the set of all matrices of rank $r$ or less. Theorem \ref{theo-matrix-rand} addresses the random  measurements case. By combining the results above, we can prove much more general results.

\begin{theo}  \label{manifold-complex-rand}
    Let $\M$ and $\Delta\M$   be a projective varieties in $\C^{p\times q}$ . Let $\A=(A_j) \in (\C^{p\times q})^N$ be randomly chosen under an absolutely continuous probability distribution in $(\C^{p\times q})^N$. Then with probability one $\A$ has the almost everywhere $\M$-recovery property if $N>\dim(\M)$, and it has the   $\M$-recovery property if $N \geq \dim(\Delta \M)$ .
\end{theo}
\Proof  We apply Theorem \ref{matrix-complex-generic} with $V_j = \C^{p\times q}$. Proposition \ref{prop:adm} implies that  the admissibility condition in the theorem is  met by any $\M$. Thus a generic $\A$ has  the almost everywhere $\M$-recovery property if $N>\dim(\M)$, and it has the   $\M$-recovery property if $N \geq \dim(\Delta \M)$ . This  implies the theorem.
\eproof

For a real projective variety $\M$, recall that it has a lift $\bar \M$ into the complex space as defined in the previous section. We have the following theorem.

\begin{theo}  \label{manifold-real-rand}
    Let $\M$ and $\Delta \M$  be a projective varieties in $\R^{p\times q}$. Let $\A=(A_j)_{j=1}^N \in (\R^{p\times q})^N$ be randomly chosen under an absolutely continuous probability distribution in $(\R^{p\times q})^N$. Then $\A$ has the almost everywhere $\M$-recovery property if $N>\dim_\R(\M)$ and $\dim(\bar\M) =\dim_\R(\M)$. It has the   $\M$-recovery property if $N \geq \dim_\R(\Delta \M)$ and  $\dim(\Delta \bar\M) =\dim_\R(\Delta \M)$.
\end{theo}
\Proof  We apply Theorem \ref{matrix-real-generic} with $V_j = \R^{p\times q}$. According to Proposition \ref{prop:adm}, the admissibility condition in the theorem are  met by $\M$. Thus a generic $\A$ has  the almost everywhere $\M$-recovery property if $N>\dim_\R(\M)$, and it has the   $\M$-recovery property if $N \geq \dim_\R(\Delta \M)$ . This  implies the theorem.
\eproof

\vspace{2mm}

\begin{coro}  \label{coro:complex-generic}
    Let $\M$   be a projective variety  in $\C^{p\times q}$. Let $\A=(A_j)_{j=1}^N \in (\C^{p\times q})^N$ where $A_j\in V_j$ is a generic element and each $V_j$ is one of the projective varieties in (A)-(C) of Proposition \ref{prop:adm}.  Then $\A$ has the almost everywhere $\M$-recovery property if $N>\dim(\M)$, and it has the   $\M$-recovery property if $N \geq \dim(\Delta \M)$ .
\end{coro}
\Proof  All the arguments for Theorem \ref{manifold-complex-rand} are valid, and the result readily follows.
\eproof

 \begin{coro}  \label{coro:real-generic}
    Let $\M$   be a projective variety  in $\R^{p\times q}$. Let $\A=(A_j)_{j=1}^N \in (\R^{p\times q})^N$ where $A_j\in V_j$ is a generic element and $\bar V_j$ for each $j$ is one of the projective varieties in (A)-(C) of Proposition \ref{prop:adm}. Then $\A$ has the almost everywhere $\M$-recovery property if $N>\dim_\R(\M)$ and $\dim(\bar\M) =\dim_\R(\M)$. It has the   $\M$-recovery property if $N \geq \dim_\R(\Delta \M)$ and  $\dim(\Delta \bar\M) =\dim_\R(\Delta \M)$.
\end{coro}
\Proof  Again, all the arguments for Theorem \ref{manifold-real-rand} are valid, and the result readily follows.
\eproof

\begin{remark}
Recall that
\[
  {\mathcal M}_{p\times q,r}(\H):=\Bigl\{Q\in \H^{p\times q}: {\rank}(Q)\leq r\Bigr\},
  \mhsp \H=\R~{\rm or}~\C.
\]
If we take $\M={\mathcal M}_{p\times q,r}(\R)$, then $\Delta \M$ is a projective variety.
For $\M={\mathcal M}_{p\times q,r}(\R)$, we also have $\dim(\bar\M) =\dim_\R(\M)=(p+q)r-r^2$ \cite[Prop. 12.2]{alge}.
Note that Theorem \ref{theo-matrix-rand}  follows from
Theorem \ref{manifold-complex-rand} while Theorem  \ref{theo-GenericMatrix} follows from  Corollary \ref{coro:complex-generic}, Corollary \ref{coro:real-generic} and Theorem \ref{matrix-complex-generic}.
 Because the admissibility condition holds for many classes of varieties, the almost everywhere rank $r$ matrix recovery property will hold in far greater generalities. It should also be pointed out that if we replace the ``generic'' stipulation in our results by random choices over some absolute continuous probability distribution over the varieties then the almost everywhere rank $r$ matrix recovery property holds with probability 1.
\end{remark}
\subsection{Rank one measurements}
The recovery of Hermitian low rank matrices from rank one measurements has attracted much attention recently \cite{ROP1,ROP2, ROP3}. In this topic, one is interested in recovering a Hermitian low rank matrix $P\in \C^{p\times p}$ from  $ \{ \vx_j^* P \vx_j \}_{j=1}^N $ where $\vx_j\in \C^p$ (For the real case, $P$ is assumed to be symmetric). One already develops many algorithms to compute it. Here, we focus on the theoretical sides. Particularly, we are interested in the following question: {\em how many measurements are needed to recover Hermitian low rank matrices from the rank one measurements}.

 Although Hermitian matrices are complex, they do not form a complex variety. Thus the theorems we have here on complex recovery cannot be applied directly to the recovery of Hermitian matrices. However, they can be formulated as the affine image of a real projective variety, and from which our theorems can be applied.

In the following lemma, we present the real dimension of the set of symmetric/Hermitian matrices of rank at most $r$ which are from \cite{alge} and \cite{Herm}, respectively.


\begin{lem}(\cite{alge, Herm}) \label{symmetric-hermitian}
\begin{itemize}
    \item[ \rm (A) ]~~ Let $ \M \subset \R^{p \times p} $ be the set of all real symmetric matrices of rank at most $ r $. Then $ \M $ is a real projective variety of dimension $ pr - r(r-1)/2 $.

    \item[ \rm (B) ]~~ Let $ \M \subset \C^{p \times p} $ be the set of all Hermitian matrices of rank at most $ r $. Then $ \M $ is a real projective variety of dimension $ 2pr - r^2 $.

\end{itemize}

\end{lem}

Now we are ready to present the following theorem:
\begin{theo}  \label{hermitian-generic}
\begin{itemize}
    \item[ \rm (A) ]~~ Let $ \M \subset \R^{p \times p} $ be the set of all real symmetric matrices of rank at most $ r $ where $ r \leq p/2 $. Let $ \vx_1, \vx_2, \dots, \vx_N $ be randomly chosen vectors in $ \R^p $ according to some absolutely continuous probability distribution. Then any $ P \in \M $ can be recovered with probability one from $ \{ \vx_j^T P \vx_j \}_{j=1}^N $ if $ N \geq 2pr-2r^2+r $, and almost all $ P \in \M $  can be recovered with probability one from $ \{ \vx_j^T P \vx_j \}_{j=1}^N $ if $ N \geq pr - r(r-1)/2 + 1$.

    \item[ \rm (B) ]~~ Let $ \M \subset \C^{p \times p} $ be the set of all Hermitian matrices of rank at most $ r $ where $ r \leq p/2 $. Let $ \vx_1, \vx_2, \dots, \vx_N $ be randomly chosen vectors in $ \C^p $ according to some absolutely continuous probability distribution. Then from $ \{ \vx_j^* P \vx_j \}_{j=1}^N $, any $ P \in \M $ can be recovered with probability one if $ N \geq 4pr-4r^2 $, and almost all $ P \in \M $  can be recovered with probability one if $ N \geq 2pr - r^2+1 $.

\end{itemize}

\end{theo}

\Proof (A) Let $V_j\subset \R^{p\times p}$ be the set of symmetric matrices of rank at most 1.
Note that $\vx_j^T P\vx_j=\Tr(A_jP)$ where $A_j=\vx_j\vx_j^T\in V_j$.
The admissibility condition of $V_j$ is already verified in the previous paper (see the proof of Theorem 4.1 in \cite{WaXu16}). By Lemma \ref{symmetric-hermitian}, $ \M $ is a real projective variety with $ \dim_\R(\M) = pr - r(r-1)/2 $. Observe that $ \Delta\M  $ is all the real symmetric matrices with rank at most $ 2r $, and then $ \dim_\R(\Delta\M) = 2pr - r(2r-1) $. Then by Theorem \ref{manifold-real-rand}, we only need to show $ \dim(\bar{\M}) = \dim_{\R}(\M) $ and $ \dim(\Delta\bar{\M}) = \dim_{\R}(\Delta\M ) $, where $ \bar{\M} $ is the lift of $ \M $ into complex space. \\
Let $ \bar{\M} \subset \C^{p \times p} $ be the set of all complex symmetric matrices of rank at most $ r $. $ \bar{\M} $ is complex projective variety and $ \bar{\M} \cap \R^{p \times p} = \M $. \\
Then using the same approach as in the proof of Lemma \ref{symmetric-hermitian}, replacing $ \R $ with $ \C $, we have $ \dim(\bar{\M}) = 2pr - r(2r-1) $. Thus $ \dim(\bar{\M}) = \dim_{\R}(\M) $. Similarly, we can show $ \dim(\Delta\bar{\M}) = \dim_{\R}(\Delta\M) $. \\
For (B), consider the map $ \varphi: \C^{p \times p} \to \C^{p \times p} $ defined by
$$
\varphi(A) = \frac{1}{2} (A + A^T) + \frac{i}{2} (A - A^T).
$$
Then $ \varphi $ is a isomorphism on $ \C^{p \times p} $.
Let
$$
\bar{\NN} = \{ A \in \C^{p \times p}: \rank(\varphi(A)) \leq r \}
$$
and
$$
\NN = \{ A \in \R^{p \times p}: \rank(\varphi(A)) \leq r \}.
$$
Then $ \bar{\NN} \cap \R^{p \times p} = \NN $. Besides, we have $ \bar{\NN} = \varphi^{-1} (\{ B \in \C^{p \times p} \text{ with rank at most } r \}) $ and $ \NN = \varphi^{-1} (\M) $. We only need to show that any $ B \in \NN $ can be recovered from $ \{ \vx_j^* \varphi^{-1}(B) \vx_j \}_{j=1}^{N} $ if $ N \geq 4pr - 4r^2 $, and almost all $ B \in \NN $ can be recovered with probability one if $ N \geq 2pr - r^2 + 1 $. Let $V_j\subset \C^{p\times p}$ be the set of Hermitian matrices with rank at most 1. A simple observation is that
$\vx_j^* \varphi^{-1}(B) \vx_j=\Tr(A_j\varphi^{-1}(B))$ where $A_j=\vx_j\vx_j^*\in V_j$.
Recall that the Hermitian matrix set $V_j$ satisfies the admissibility condition (see Theorem 4.1 in \cite{WaXu16}).
The admissibility condition naturally holds since $ \varphi $ is a linear transformation on $ \C^{p \times p} $. According to Lemma \ref{symmetric-hermitian}, $\dim_\R(\NN)=\dim_\R(\M)=2pr-r^2$. Since $\varphi$ is a linear transform, we have $\dim(\bar \NN)=2pr-r^2$. Hence, $\dim_\R(\NN)=\dim(\bar \NN)$. The conclusion follows from Theorem \ref{matrix-real-generic}.
\eproof

\end{document}